\documentclass[prl,superscriptaddress,twocolumn]{revtex4}

\usepackage{amsfonts,amssymb,amsmath}
\usepackage[]{graphics,graphicx,epsfig}
\usepackage{amsthm}

\def\identity{\leavevmode\hbox{\small1\kern-3.8pt\normalsize1}}

\renewcommand{\epsilon}{\varepsilon}

\begin{document}

%\preprint{APS/123-QED}

\title{Quantum walk as a generalized measuring device}

\author{Pawe{\l} Kurzy\'nski}
\email{cqtpkk@nus.edu.sg}
\affiliation{Centre for Quantum Technologies,
National University of Singapore, 3 Science Drive 2, 117543 Singapore,
Singapore}
\affiliation{Faculty of Physics, Adam Mickiewicz University, Umultowska
85, 61-614 Pozna\'{n}, Poland.}

\author{Antoni W{\'o}jcik}%
\affiliation{Faculty of Physics, Adam Mickiewicz University, Umultowska
85, 61-614 Pozna\'{n}, Poland.}

\date{\today}% It is always \today, today,

\begin{abstract}
We show that a one-dimensional discrete time quantum walk can be used to implement a generalized measurement in terms of positive operator value measure (POVM) on a single qubit. More precisely, we show that for a single qubit any set of rank 1 and rank 2 POVM elements can be generated by a properly engineered quantum walk. In such a scenario the measurement of particle at position $x=i$ corresponds to a measurement of a POVM element $E_i$ on a qubit.  We explicitly construct quantum walks implementing unambiguous state discrimination and SIC-POVM. 
\end{abstract}

%\pacs{}% PACS, the Physics and Astronomy
\maketitle

{\it Introduction.} Discrete time quantum walk is a process in which the evolution of a quantum particle on a lattice depends on a state of an auxiliary system (coin). In the simplest version of a one-dimensional quantum walk the coin is a two-level system. The particle moves either one step to the left or to the right depending on the state of the coin.  Between subsequent steps the state of the coin evolves, which after many steps of the walk results in nontrivial correlations between the coin and the particle's position and in a spatial probability distribution that in general cannot be reproduced by classical random walks \cite{ADZ,rew,ent}. 

A single step of an initially localized quantum walk can be considered as a projective von Neumann measurement of the coin, because if one finds the particle at position $x \pm 1$ one knows that the state of the coin corresponds to the "right/left shift". However, in general one can allow the system to evolve for more than one step before the position measurement is done. In this case the particle can be found in more than two different positions and one may ask whether the measurement of particle at position $x=i$ corresponds to some generalized measurement of the qubit, i.e., a positive operator value measure (POVM) on the coin state. In this work we investigate such a possibility. 

POVMs allow one to gain more information from a single measurement than the standard von Neumann projective measurements. Their wide applicability include discrimination of quantum states \cite{statediscr} and quantum state tomography in terms of symmetric informationally complete (SIC) POVMs \cite{SIC}. Physically, POVMs correspond to projective measurements on a joint system of the system of interest and an ancilla whose state is known. Mathematically, POVM elements $E_i$ are given by $E_i=\text{Tr}_{anc}\{(\openone\otimes\sigma)\pi_i\}$, where $\openone$ is the identity operator on the Hilbert space of the system, $\sigma$ the state of an ancilla, $\pi_i$ is the von Neumann projector on the joint Hilbert space and one traces out an ancillary system. The probability of measuring the $i$'th POVM element on a state $\rho$ is given by $\text{Tr}\{E_i \rho\}$. In addition to the non-negativity condition $E_i \geq 0$ the complete set of measurement operators has to have the resolution of identity, therefore POVM elements obey 
$\sum_i E_i=\openone$. In case of quantum walks the role of ancilla is played by the position of the particle and the form of the effective projector $\pi_i$ stems from the nontrivial evolution of the walk.

For a one-dimensional discrete time quantum walk the state of the system is described by two degrees of freedom $|x,c\rangle$, the position of the particle $x=\dots,-1,0,1,\dots$ and the coin $c=\rightarrow,\leftarrow$. Since the dynamics of the system is discrete, one step is given by the unitary operator $U(t,t+1)=TC(x,t)$, where 
\begin{equation}
T=\sum_x|x+1,\rightarrow\rangle\langle x,\rightarrow|+|x-1,\leftarrow\rangle\langle x,\leftarrow| \nonumber
\end{equation}
is the conditional translation operator and $C(x,t)$ is a coin operator whose action in general can depend on position and time 
\begin{eqnarray}
C(x,t)|x,\rightarrow\rangle &=& \alpha(x,t)|x,\rightarrow\rangle+\beta(x,t)|x,\leftarrow\rangle, \nonumber \\ C(x,t)|x,\leftarrow\rangle &=& \beta^{\ast}(x,t)|x,\rightarrow\rangle-\alpha^{\ast}(x,t)|x \leftarrow\rangle. \nonumber
\end{eqnarray} 
Throughout the paper we use the notation $|\rightarrow\rangle=(1, 0)^T$ and $|\leftarrow\rangle=(0,1)^T$.

The one step is analogical to an action of Stern-Gerlach device (S-G) on spin $1/2$ particles. The action of the coin operator corresponds to spin rotation which is analogical to the choice of a specific direction of S-G or an application of a linear magnetic field, whereas the splitting of the particle beam is obtained via an application of $T$, which corresponds to a magnetic field gradient. In principle, before the actual measurement happens (particles hit the screen) one can perform another spin rotation (coin operation) and beam splitting ($T$ operation). Interesting effects may happen if one allows different particle beams to interfere before they hit the screen.

Quantum walks are computationally more efficient than their classical counterparts and were shown to be able to efficiently solve a number of problems \cite{Amb}. In particular, it was shown that they are capable of universal quantum computation \cite{comp}. Moreover, quantum walks were implemented in laboratory and it was shown that one can achieve a substantial control over the evolution of a quantum walker \cite{lab}. It is therefore of great importance to investigate and to exploit all the possibilities that quantum walks can offer.   

The rest of the paper is organized as follows. We provide a simple example of a quantum walk that is capable of optimal unambiguous state discrimination of two equally probable qubit quantum states. The analysis of this example allows us to construct a general algorithm for a generation of arbitrary POVM elements. We then discuss how to modify a standard S-G apparatus, which in many textbooks is considered as a flagship example of von Neumann measurement, to allow it to perform generalized measurements. The modification is based on our quantum walk model. In particular, we consider implementation of the state discrimination protocol. Derivation of POVM elements for the state discrimination protocol and an explicit quantum walk algorithm for generation of SIC POVM are given in the appendix.

{\it Example: unambiguous state discrimination.} Imagine that one is given one of two qubit states $\{|0\rangle,\alpha|0\rangle+\beta|1\rangle\}$ with equal a priori probabilities. The goal is to find which of the two states was given. We consider three possible answers to this test: {\it it is definitely state 1}, {\it it is definitely state 2}, or {\it I do not know}. 

Now, let us introduce a quantum walk capable of achieving this goal. First, we note that it is always possible to represent the two states as $|\psi_{\pm}\rangle=\cos\frac{\theta}{2}|\rightarrow\rangle \pm \sin\frac{\theta}{2}|\leftarrow\rangle,$ where $\theta\in [0,\frac{\pi}{2}]$. For the first step we choose the coin operator to be trivial $C(x,0)=\openone$, hence the state after one QW step is $$|\psi_{\pm}(1)\rangle=\cos\frac{\theta}{2}|1,\rightarrow\rangle\pm\sin\frac{\theta}{2}|-1,\leftarrow\rangle.$$ For the next step the coin operators are $$C(-1,1)=NOT=\begin{pmatrix} 0 & 1 \\ 1 & 0 \end{pmatrix},$$ $$C(1,1)=\begin{pmatrix} \sqrt{1-\tan^2\frac{\theta}{2}} & \tan\frac{\theta}{2} \\ \tan\frac{\theta}{2} & -\sqrt{1-\tan^2\frac{\theta}{2}} \end{pmatrix},$$ and identity elsewhere, therefore the state after coin operation is
$$C(x,1)|\psi_{\pm}(1)\rangle=\sqrt{\cos\theta}|1,\rightarrow\rangle+\sin\frac{\theta}{2}|1,\leftarrow\rangle\pm\sin\frac{\theta}{2}|-1,\rightarrow\rangle$$ and after translation
$$|\psi_{\pm}(2)\rangle=\sqrt{\cos\theta}|2,\rightarrow\rangle+\sin\frac{\theta}{2}|0,\leftarrow\rangle\pm\sin\frac{\theta}{2}|0,\rightarrow\rangle.$$ Finally, for the third step the coin operator is identity everywhere except for position $x=0$ for which it is the Hadamard operator $$C(0,2)=H=\frac{1}{\sqrt{2}}\begin{pmatrix} 1 & 1 \\ 1 & -1 \end{pmatrix},$$ therefore $$|\psi_{\pm}(3)\rangle=\sqrt{\cos\theta}|3,\rightarrow\rangle\pm\sqrt{2}\sin\frac{\theta}{2}|\pm1,\rightarrow\rangle.$$ As a consequence, if one measures particle at position $x=1$, one immediately knows that the coin was in $|\psi_+\rangle$ state, if at position $x=-1$ it was $|\psi_-\rangle$ state, and if at position $x=3$ one learns nothing. In the appendix A we show that the above quantum walk indeed generates the POVM elements corresponding to the unambiguous state discrimination problem.

{\it Generation of arbitrary rank 1 POVM elements.} We focus on rank 1 POVMs, since higher rank POVMs can be constructed as a convex combination of rank 1 elements. We will come back to this problem at the end of the next section. In case of a single qubit the orthogonality relation between states and rank 1 POVM elements uniquely determines the latter up to the magnitude. This property is going to be our main tool in the construction of a quantum walk algorithm for POVM elements. 

To clarify what we mean, let us start with the following observation. Imagine that the initial state of the quantum walk is localized at position $x=0$ and that the initial coin state is either one of the two orthogonal states $|\psi\rangle$ or $|\psi_{\perp}\rangle$. Next, assume that at some stage of the quantum walk there is nonzero probability of finding the particle at position $x$ for both initial states. If we apply a unitary coin operation at position $x$ such that after the subsequent translation there is nonzero probability of finding the particle at position $x+1$ for the initial state $|\psi\rangle$, but the probability of finding the particle at position $x+1$ is zero for the initial state $|\psi_{\perp}\rangle$, then the measurement corresponding to position $x$ has to correspond to rank 1 POVM element of the form $a|\psi\rangle\langle\psi|$, where $0<a\leq 1$. Here we assume that the probability that particle is at position $x+1$ could only be due to translation from $x$, not $x+2$ (due to a reason that will become clear in a moment). Furthermore, note that it is always possible to find a proper coin operator. The coin states at position $x$ corresponding to initial coin states $|\psi\rangle$ and $|\psi_{\perp}\rangle$ can be denoted as $|\phi\rangle$ and $|\phi'\rangle$, respectively. These two states do not have to be orthogonal. It is clear that there always exists a unitary operation that transforms $|\phi'\rangle$ into $|\leftarrow\rangle$, therefore if we apply this operation as a coin transformation at position $x$, the subsequent translation will be always to position $x-1$, not $x+1$, given the initial coin state was $|\psi_{\perp}\rangle$. 

Let us now propose an algorithm for the generation of an arbitrary rank 1 POVM $\{E_1,\dots,E_n\}$, where $E_i = a_i|\psi_i\rangle\langle\psi_i|$. 
\begin{enumerate}
\item Initiate the quantum walk at position $x=0$ with the coin state corresponding to the qubit state one wants to measure 
\item Set $i := 1$
\item While $i < n$ do:
\begin{enumerate}
\item Apply coin operation $C_i^{(1)}$ at position $x=0$ and identity elsewhere and then apply translation operator $T$
\item Apply coin operation $C_i^{(2)}$ at position $x=1$, $NOT$ at position $x=-1$ and identity elsewhere and then apply translation operator $T$
\item $i:=i +1$
\end{enumerate}
\end{enumerate}

The POVM that is going to be generated depends solely on the coin operators $C_i^{(1)}$ and $C_i^{(2)}$ which are of the form
\begin{equation}\label{c1}
C_i^{(1)}=|\rightarrow\rangle\langle \phi'_{i\perp}|+|\leftarrow\rangle\langle \phi'_i|
\end{equation}
and
\begin{equation}\label{c2}
C_i^{(2)}=\begin{pmatrix} \cos\theta_i & \sin\theta_i \\ \sin\theta_i & -\cos\theta_i \end{pmatrix}.
\end{equation}
In the above $|\phi'_i\rangle$ is the state that at this stage of the algorithm would correspond to the coin state at position $x=0$ if the initial coin state was $|\psi_{i\perp}\rangle$ ($\langle \psi_i|\psi_{i\perp}\rangle=0$) and $|\phi'_{i\perp}\rangle$ is a state orthogonal to $|\phi'_i\rangle$. This guarantees that after the step 3(a) the POVM element at position $x=1$ is proportional to $|\psi_i\rangle\langle\psi_i|$.  The angle $\theta_i$ is chosen such that after the step 3(b) the element at position $x=2$ has proper magnitude $a_i$.

Before we prove the algorithm's capability of generating an arbitrary rank 1 POVM, let us note that it generates spatial probability distribution over positions $x=2k$, $k=0,\dots,n-1$. The POVM element $E_i$ corresponds to the measurement of particle at position $x=2(n-i)$. Note, that the previously considered algorithm for unambiguous state discrimination is in fact a special case of the above one (without the last step 3b). The interference phenomenon occurs only at position $x=0$ and once the particle passes the position $x=1$ it can only move to the right (which clarifies our previous assumption that there is no translation from $x+2$ to $x+1$). On the other hand, due to the $NOT$ operation at position $x=-1$ the particle can never pass beyond this point.

{\it Proof of universality.} We have already shown that a proper choice of $C_i^{(1)}$ allows one to generate an element proportional to $a_i|\psi_i\rangle\langle\psi_i|$ for any $|\psi_i\rangle$. What is left to show is that the algorithm gives one a possibility to manipulate with the magnitude $a_i$. We follow the inductive approach. Firstly, note that for the first repetition of the subroutine 3(a)-(c) ($i=1$) the state of the particle is initially localized at $x=0$ and the step 3(a) corresponds to a standard von Neumann measurement. In particular, measurement of particle at position $x=1$ and $x=-1$ corresponds to qubit projections onto $|\psi_1\rangle\langle \psi_1|$ and $|\psi_{1\perp}\rangle\langle \psi_{1\perp}|$, respectively. The subsequent step 3(b) results in two POVM elements. The first one $\cos^2\theta_1|\psi_1\rangle\langle \psi_1|$ corresponding to $x=2$ and the second one $|\psi_{1\perp}\rangle\langle \psi_{1\perp}|+\sin^2\theta_1|\psi_1\rangle\langle \psi_1|$ corresponding to $x=0$. Note that $\theta_1$ can be chosen such that $\cos^2\theta_1=a_1$, therefore the element at position $x=2$ corresponds to $E_1$ and the one at $x=0$ to $\openone - E_1$.

Next, let us assume that the quantum walk has already generated elements $\{E_1,E_2\dots,E_k\}$  corresponding to positions $x=2k,2(k-1),\dots,2,$ respectively ($k<n-1$ since the generation of $E_{n-1}$ implies also the generation of $E_{n}$ due to the fact that POVM elements have the resolution of identity). The resolution of identity implies that the measurement at position $x=0$ corresponds to ${\cal E}=\openone - \sum_{j=1}^k E_j$. We apply $C_{k+1}^{(1)}$ such that after the step 3(a) the element ${\cal E}$ splits into $E'_{k+1}=a|\psi_{k+1}\rangle\langle\psi_{k+1}|$ at position $x=1$ and ${\cal E} - E'_{k+1}$ at position $x=-1$. After the step 3(b) the element at position $x=2$  is $a\cos^2\theta_{k+1}|\psi_{k+1}\rangle\langle\psi_{k+1}|$, which corresponds to $E_{k+1}$ if $a\cos^2\theta_{k+1}=a_{k+1}$. In general the range of $a_{k+1}$ can be $0\leq a_{k+1} \leq a_{\max}$, where $a_{max}$ arises from the non-negativity constraint ${\cal E} - a_{max}|\psi_{k+1}\rangle\langle\psi_{k+1}|\geq 0$. Below we show that $a=a_{max}$, therefore it is always possible to choose an appropriate angle $\theta_{k+1}$ such that $E_{k+1}$ is generated at position $x=2$. 

%A schematic picture presenting the steps 3(a) and 3(b) discussed in this proof is given in the Fig.1.
%\begin{figure}[t]
%\vspace{-2cm}
%\begin{center}
%\includegraphics[scale=0.6]{fig1.pdf}
%\end{center}
%\vspace{-10.5cm}
%\caption{Schematic picture presenting the steps 3(a) and 3(b) discussed in the proof of universality.}
%\end{figure}

Since POVM elements act in a two-dimensional space the element ${\cal E} - a_{max}|\psi_{k+1}\rangle\langle\psi_{k+1}|$ is a unique rank 1 operator because at least one of its eigenvalues is zero. Therefore, the goal is to show that after the step 3(a) the element corresponding to position $x=-1$, given by ${\cal E} - a|\psi_{k+1}\rangle\langle\psi_{k+1}|$, is also a rank 1 operator. 
This is trivial, since the measurement of the quantum walker at position $x=-1$ gives us full information about the coin state, namely that it is $|\leftarrow\rangle$, which can only happen if the measurement operator is rank 1. 

%Note, that for a rank 1 qubit POVM element there always exists a unique state that is orthogonal to this element. On the other hand, if ${\cal E} - a|\psi_{k+1}\rangle\langle\psi_{k+1}|$ is rank 2 element then there is no such state. Let us assume that this POVM element is of rank 2. If we set $\theta_{k+1}=0$ then for the step 3(b) there is no translation from $x=1$ to $x=0$, therefore the element at $x=0$ is again ${\cal E} - a|\psi_{k+1}\rangle\langle\psi_{k+1}|$. Now, if for the next step 3(a) ($i=k+2$) we chose $C_{k+2}^{(1)}=NOT$ there would be no translation from $x=0$ to $x=1$, however this would imply that there exists a state orthogonal to ${\cal E} - a|\psi_{k+1}\rangle\langle\psi_{k+1}|$ which contradicts our assumption that it is the rank 2 element. This ends the proof.

Finally, let us discuss the generation of rank 2 POVM elements. These elements can be constructed as a convex combination of two rank 1 elements. The above algorithm can be modified in the following way. Imagine that the algorithm generated $N$ POVM elements corresponding to the measurement of the quantum walker at positions $x=0,2,4,\dots,2(N-1)$ and that we want to construct a rank 2 element that is a combination of two rank 1 elements corresponding to positions $x=2i$ and $x=2j$ ($j>i$). It is enough to apply $NOT$ coin operation at position $x=2j$ and then to apply conditional translation $j-i$ times. As a result, the two probability amplitudes originating from positions $x=2i$ and $x=2j$ meet at position $x=i+j$ and the respective position measurement corresponds to the POVM element $E_{i+j}=E_i + E_j$. The other rank 1 POVM elements are also shifted to the right by $j-i$ positions, however their structure remains unchanged.   

{\it Generalization of S-G apparatus.} The main physical mechanism that allows quantum walks to implement POVM is the interference between the probability amplitudes of the quantum walker. This interference is not present in the standard von Neumann model of measurement (see for example \cite{AR}), whose flagship example is the S-G apparatus. However, in principle one may modify this apparatus to incorporate interference between particle beams.

Let us consider a beam of spin-1/2 particles incident upon S-G apparatus. Due to the gradient of magnetic field the beam splits into two beams. The upper beam contains particles whose spin is oriented "up" along the respective direction, and the lower beam contains particles whose spin is oriented "down". If we placed a screen behind the S-G magnet we would perform a standard projective measurement in the "up"/"down" basis. However, we may continue the experiment by allowing two beams to pass through different homogeneous linear magnetic fields that would cause different rotations of spins. These beams fan out, but if we use additional S-G magnets we may split them again and some of new beams would overlap. If this overlap happens in another S-G magnet one may cause the beams to interfere (just like in an optical beam splitter).

\begin{figure}[t]
\begin{center}
\includegraphics[scale=0.45]{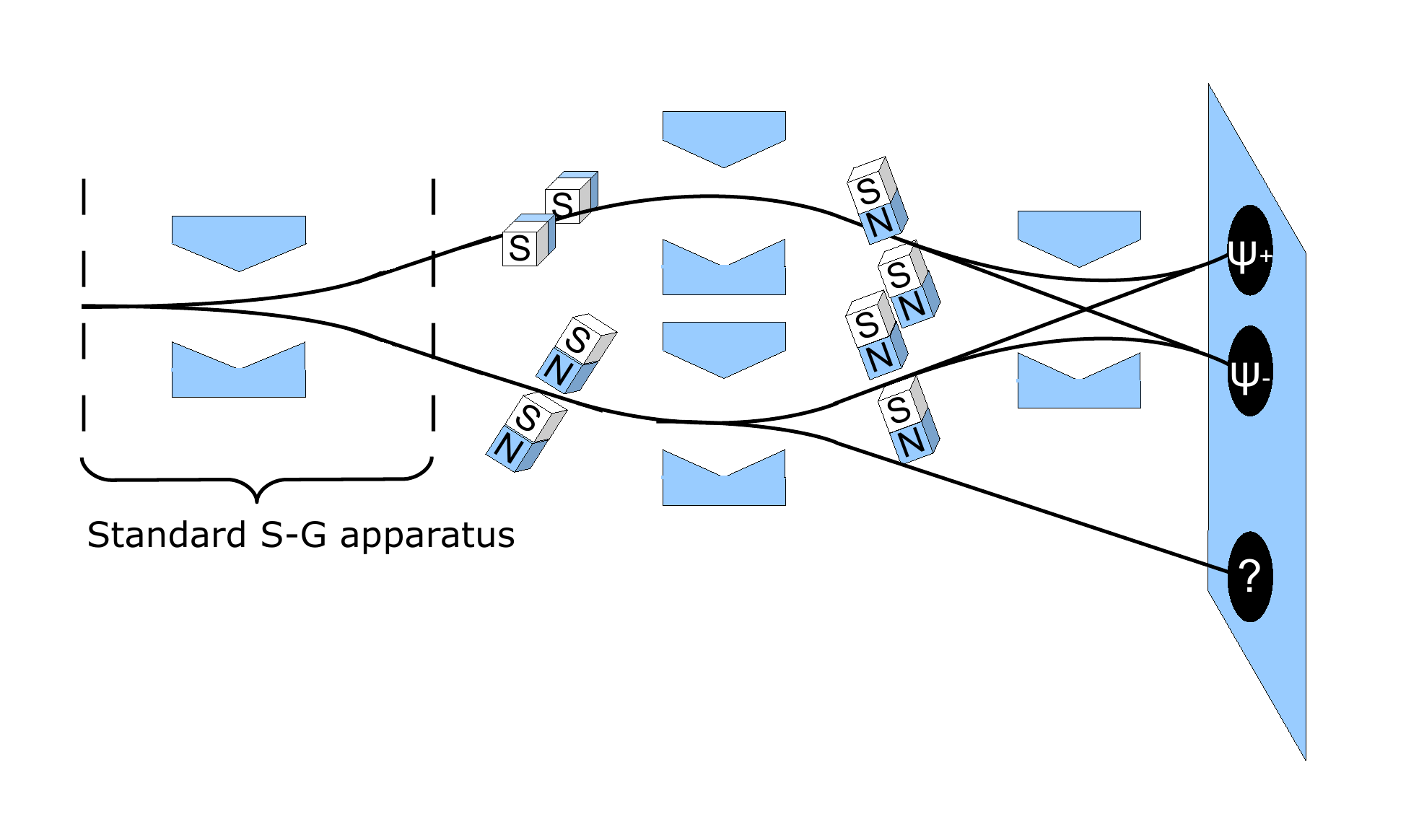}
\end{center}
\vspace{-10mm}
\caption{\label{f1} Schematic picture representing generalization of S-G apparatus that incorporates POVM. In particular, the above setting visualizes the unambiguous state discrimination scenario. The heart of the setting is the last S-G magnet whose magnetic field gradient allows for an interference effect between two spin-1/2 particle beams. Block magnets in between S-G magnets represent action of different homogeneous linear magnetic fields on the beams and correspond to coin operations. S-G magnets correspond to conditional translations.}
\end{figure}
 
In Fig. \ref{f1} we sketch a S-G implementation of the unambiguous state discrimination protocol based on the quantum walk that we introduced in the beginning of this work. The particle beam is split into two parts that pass through different homogeneous linear magnetic fields. In particular, the spin orientation of the upper beam is changed from "up" to "down" by the $\pi$ rotation about an axis perpendicular to the beam. Due to this fact, the subsequent S-G magnet does not split this beam, but rather changes the direction of its propagation. As a result, after the last S-G magnet there are only three beams corresponding to the outcomes: {\it it is definitely state $\psi_+$}, {\it it is definitely state $\psi_-$}, and {\it I do not know}. 

{\it Conclusions.} We proved that discrete-time quantum walks are capable of performing generalized measurements on a single qubit. The main physical effect employed in this process is the interference between the probability amplitudes of the quantum walker. In particular, we proposed a quantum walk algorithm for the generation of arbitrary rank 1 and rank 2 POVM elements for a single qubit. Moreover, we explicitly constructed quantum walks for unambiguous state discrimination of two equally probable qubit states and SIC-POVM (see appendix B). Finally, using quantum walk approach we considered a generalization of S-G apparatus. 

We acknowledge stimulating discussions with Ravishankar Ramanathan.  This work is supported by the National Research Foundation and Ministry of Education in Singapore. P. K. is also supported by the Foundation for Polish Science.

\section{Appendix A}

Here we the quantum walk generation of the POVM elements corresponding to the unambiguous state discrimination problem. We are going to consider only one POVM element, since the construction of the other two follows from this example. The initial state of the ancilla (position) is $\sigma=|x=0\rangle\langle x=0|$, whereas the projector $\pi_i$ corresponds to $\pi_i=U^{\dagger} \left( |x=i\rangle\langle x=i| \otimes \openone \right) U$. The unitary operator $U$ generates the three steps of the above quantum walk, and the identity operator acts on the coin space. In order to evaluate the form of the projector $\pi_i$ one has to consider the reversed quantum walk evolution due to $U^{\dagger}$ on both states $|i,\rightarrow\rangle$ and $|i,\leftarrow\rangle$. Finally, in order to obtain the POVM element $E_i$ one has to consider the overlap of $\pi_i$ with the ancilla state and then trace over ancilla.

Let us consider the POVM element $E_{-1}$, i.e., the element corresponding to finding the particle at position $x=-1$. We start with the state $|-1,\leftarrow\rangle$. The first step of the reversed evolution corresponds to $C(x,2)^{\dagger}T^{\dagger}$. The reversed translation results in the state $|0,\leftarrow\rangle$ and the application of the coin operator gives $\frac{1}{\sqrt{2}}\left(|0,\rightarrow\rangle - |0,\leftarrow\rangle\right)$. For the second step, the translation gives  $\frac{1}{\sqrt{2}}\left(|-1,\rightarrow\rangle - |1,\leftarrow\rangle\right)$ and the coin operation gives   $\frac{1}{\sqrt{2}}\left(|-1,\leftarrow\rangle - \tan\frac{\theta}{2}|1,\rightarrow\rangle + \sqrt{1-\tan^2\frac{\theta}{2}}|1,\leftarrow\rangle\right)$. Finally, the last step is just reversed translation which results in $\frac{1}{\sqrt{2}}\left(|0,\leftarrow\rangle - \tan\frac{\theta}{2}|0,\rightarrow\rangle + \sqrt{1-\tan^2\frac{\theta}{2}}|2,\leftarrow\rangle\right)$. Taking the overlap with the ancilla state $|x=0\rangle$ one finds that the contribution of the above state to the POVM is $\frac{1}{\sqrt{2}}\left(|\leftarrow\rangle - \tan\frac{\theta}{2}|\rightarrow\rangle\right)$.  

On the other hand, it is easy to see that the state $|-1,\rightarrow\rangle$ does not contribute to the POVM element, since the reverse evolution generates a state $|-4,\rightarrow\rangle$ that has no overlap with position $x=0$. Therefore, the POVM element is given by
\begin{eqnarray}
E_{-1} & = & \frac{1}{2 \cos^2\frac{\theta}{2}}\left(\cos\frac{\theta}{2}|\leftarrow\rangle - \sin\frac{\theta}{2}|\rightarrow\rangle\right) \nonumber \\ & \times & \left(\cos\frac{\theta}{2}\langle\leftarrow| - \sin\frac{\theta}{2}\langle\rightarrow|\right)\nonumber
\end{eqnarray}
which is the correct POVM element for unambiguous state discrimination since $E_{-1}|\psi_+\rangle = 0$ \cite{statediscr}.

\section{Appendix B}

In this section we introduce a quantum walk that generates symmetric informationally complete (SIC) POVM \cite{SIC}. The POVM elements are proportional to projectors onto the following qubit (coin) states:
\begin{eqnarray}
|\psi_1\rangle & = & |\rightarrow\rangle, \nonumber \\
|\psi_2\rangle & = & \sqrt{\frac{1}{3}}|\rightarrow\rangle + \sqrt{\frac{2}{3}}|\leftarrow\rangle, \nonumber \\
|\psi_3\rangle & = & \sqrt{\frac{1}{3}}|\rightarrow\rangle + e^{i\frac{2\pi}{3}}\sqrt{\frac{2}{3}}|\leftarrow\rangle, \nonumber \\
|\psi_4\rangle & = & \sqrt{\frac{1}{3}}|\rightarrow\rangle + e^{-i\frac{2\pi}{3}}\sqrt{\frac{2}{3}}|\leftarrow\rangle. \nonumber
\end{eqnarray}
Note, that $|\langle\psi_i|\psi_j\rangle|^2=\frac{1}{3}$ for $i\neq j$.

The coin operators for the generation of SIC-POVM (see the steps of the algorithm) are
\begin{eqnarray}
C_1^{(1)} &=& \openone,~~C_1^{(2)}= \frac{1}{\sqrt{2}} \begin{pmatrix} 1 & 1 \\ 1 & -1  \end{pmatrix}, \nonumber \\
C_2^{(1)} &=& \frac{1}{\sqrt{2}} \begin{pmatrix} -1 & 1 \\ 1 & 1  \end{pmatrix},~~C_2^{(2)}= \frac{1}{\sqrt{3}} \begin{pmatrix} \sqrt{2} & 1 \\ 1 & -\sqrt{2}  \end{pmatrix}, \nonumber \\
C_3^{(1)} &=& \frac{1}{\sqrt{2}} \begin{pmatrix} e^{-i\frac{\pi}{3}} & e^{i\frac{\pi}{6}} \\ e^{i\frac{\pi}{3}} & e^{-i\frac{\pi}{6}}  \end{pmatrix},~~C_3^{(2)}= \openone. \nonumber 
\end{eqnarray}

To visualize the action of the above coin operators, let us consider the algorithm for the four states 
\begin{eqnarray}
|\tilde \psi_1\rangle & = & |\leftarrow\rangle, \nonumber \\
|\tilde \psi_2\rangle & = & \sqrt{\frac{2}{3}}|\rightarrow\rangle - \sqrt{\frac{1}{3}}|\leftarrow\rangle, \nonumber \\
|\tilde \psi_3\rangle & = & \sqrt{\frac{2}{3}}|\rightarrow\rangle - e^{i\frac{1\pi}{3}}\sqrt{\frac{2}{3}}|\leftarrow\rangle, \nonumber \\
|\tilde \psi_4\rangle & = & \sqrt{\frac{2}{3}}|\rightarrow\rangle - e^{-i\frac{1\pi}{3}}\sqrt{\frac{2}{3}}|\leftarrow\rangle. \nonumber
\end{eqnarray}
Observe that $\langle \tilde \psi_j|\psi_j\rangle = 0$. After the first subroutine (3 a-c for $i=1$) the states are
\begin{eqnarray}
|\tilde \psi_1 (1)\rangle & = & |0,\rightarrow\rangle, \nonumber \\
|\tilde \psi_2 (1)\rangle & = & \sqrt{\frac{1}{3}}(|2,\rightarrow\rangle - |0,\rightarrow\rangle - |0,\leftarrow\rangle), \nonumber \\
|\tilde \psi_3 (1)\rangle & = & \sqrt{\frac{1}{3}}(|2,\rightarrow\rangle - e^{i\frac{2\pi}{3}}|0,\rightarrow\rangle - |0,\leftarrow\rangle), \nonumber \\
|\tilde \psi_4 (1)\rangle & = & \sqrt{\frac{1}{3}}(|2,\rightarrow\rangle - e^{-i\frac{2\pi}{3}}|0,\rightarrow\rangle - |0,\leftarrow\rangle). \nonumber
\end{eqnarray}
After the second subroutine ($i=2$)
\begin{eqnarray}
|\tilde \psi_1 (2)\rangle & = & -\sqrt{\frac{1}{3}}|2,\rightarrow\rangle -\sqrt{\frac{1}{6}}|0,\leftarrow\rangle  + \sqrt{\frac{1}{2}}|0,\rightarrow\rangle, \nonumber \\
|\tilde \psi_2 (2)\rangle & = & \sqrt{\frac{1}{3}}|4,\rightarrow\rangle - \sqrt{\frac{2}{3}} |0,\rightarrow\rangle, \nonumber \\
|\tilde \psi_3 (2)\rangle & = & \sqrt{\frac{1}{3}}|4,\rightarrow\rangle - \sqrt{\frac{1}{3}} e^{-i\frac{\pi}{6}}|2,\rightarrow\rangle  \nonumber \\ 
&-& \sqrt{\frac{1}{6}}(e^{i\frac{\pi}{3}}|0\rightarrow\rangle+e^{-i\frac{\pi}{6}}|0,\leftarrow\rangle), \nonumber \\
|\tilde \psi_4 (2)\rangle & = & \sqrt{\frac{1}{3}}|4,\rightarrow\rangle - \sqrt{\frac{1}{3}} e^{i\frac{\pi}{6}}|2,\rightarrow\rangle \nonumber \\
&-& \sqrt{\frac{1}{6}}(e^{-i\frac{\pi}{3}}|0\rightarrow\rangle+e^{i\frac{\pi}{6}}|0,\leftarrow\rangle). \nonumber
\end{eqnarray}
Finally, after the third subroutine ($i=3$)
\begin{eqnarray}
|\tilde \psi_1 (3)\rangle & = & \sqrt{\frac{1}{3}}(-|4,\rightarrow\rangle + e^{i\frac{\pi}{3}}|2,\rightarrow\rangle  + e^{-i\frac{\pi}{3}}|0,\rightarrow\rangle), \nonumber \\
|\tilde \psi_2 (3)\rangle & = &  \sqrt{\frac{1}{3}}(|6,\rightarrow\rangle - e^{-i\frac{\pi}{3}}|2,\rightarrow\rangle  - e^{i\frac{\pi}{3}}|0,\rightarrow\rangle), \nonumber \\
|\tilde \psi_3 (3)\rangle & = & \sqrt{\frac{1}{3}}(|6,\rightarrow\rangle - e^{-i\frac{\pi}{6}}|4,\rightarrow\rangle  - |2,\rightarrow\rangle), \nonumber \\
|\tilde \psi_4 (3)\rangle & = &  \sqrt{\frac{1}{3}}(|6,\rightarrow\rangle - e^{i\frac{\pi}{6}}|4,\rightarrow\rangle  - |0,\rightarrow\rangle). \nonumber
\end{eqnarray}
It is clear that $|\tilde \psi_1 (3)\rangle$ cannot be found at position $x=6$, $|\tilde \psi_2 (3)\rangle$ at position $x=4$, $|\tilde \psi_3 (3)\rangle$ at position $x=0$ and $|\tilde \psi_4 (3)\rangle$ at position $x=2$.

\end{document}